\documentclass[10pt]{iopart}

\usepackage{graphicx}
\usepackage{subfigure}
\usepackage{verbatim}
\usepackage{dcolumn}% Align table columns on decimal point
\usepackage{bm}% bold math
\usepackage{epsf}
\usepackage{color}
\usepackage[toc]{appendix}
\usepackage{stmaryrd}
\usepackage{amsthm}
\usepackage{hhline}
\usepackage{amssymb}
\usepackage{dsfont}
\usepackage{iopams}
\usepackage{cite}
\usepackage{tikz}
\usetikzlibrary{arrows,shapes,calc,matrix}

\expandafter\let\csname equation*\endcsname\relax
\expandafter\let\csname endequation*\endcsname\relax
\usepackage{amsmath}
\usepackage{xstring}

\makeatletter
\newcommand\footnoteref[1]{\protected@xdef\@thefnmark{\ref{#1}}\@footnotemark}
\makeatother
\newcommand\myeqref[1]{Eq.\hspace{0.5mm}(\textup{\ref{#1}})}
\newcommand\invisiblesection[1]{%
	\refstepcounter{section}%
	\addcontentsline{toc}{section}{\protect\numberline{\thesection}#1}%
	\sectionmark{#1}}
\newcommand{\bra}[1]{\left\langle #1\right|}
\newcommand{\ket}[1]{\left|#1\right\rangle}

\newcommand{\trace}[1]{\mathrm{tr}\left\{#1\right\}}
\newcommand{\ptr}[2]{\mathrm{tr_{#1}}\left\{#2\right\}}

\newcommand{\la}{\left\langle}
\newcommand{\ra}{\right\rangle}
\newcommand{\pd}{\partial}

\newcommand{\miU}[1]{\min_{U}{\left\{#1\right\}}}
\newcommand{\mi}[1]{\min{\left\{#1\right\}}}

\newcommand{\ex}[1]{\exp{\left(#1\right)}}

\newcommand{\bla}{bla\\bla\\bla\\bla\\bla}
\usepackage{fmtcount}

\newcommand{\mc}[1]{\mathcal{#1}}

\newcommand{\mf}[1]{\mathfrak{#1}}

\renewcommand{\appendix}{
}

\newcommand{\draftmode}{1}    %to control draft colors below
\newcommand{\notetoself}[1]{\ifnum \draftmode=1 {\color[rgb]{0,0,0.8} [#1]} \fi}  %notes to self in blue when \draftmode==1.  invisible otherwise
\newcommand{\cuttext}[1]{\ifnum \draftmode=1 {\color[rgb]{0,0.5,0} [#1]} \fi}  %cut out text in green when \draftmode==1.  invisible otherwise
\newcommand{\warntext}[1]{\ifnum \draftmode=1 {\color[rgb]{0.9,0.6,0} #1} \else {#1} \color{black} \fi}
\newcommand{\aref}[1]{{Appendix~\hyperref[#1]{A}}}
\newcommand{\bref}[1]{{Appendix~\hyperref[#1]{B}}}
\usepackage[colorlinks=true,citecolor=blue,linkcolor=blue,urlcolor=blue]{hyperref}
\usepackage{doi}
\usepackage{cleveref}
\begin{document}

\title{Ergotropy from quantum and classical correlations}
	
\author{Akram Touil\textsuperscript{1,*}, Bar\i\c{s} \c{C}akmak\textsuperscript{2}, and Sebastian Deffner\textsuperscript{1,3}}
\address{$^1$Department of Physics, University of Maryland, Baltimore County, Baltimore, MD 21250, USA}
\address{$^2$College of Engineering and Natural Sciences, Bah\c{c}e\c{s}ehir University, Be\c{s}ikta\c{s}, \.{I}stanbul 34353, Turkey}
\address{$^3$Instituto de F\'isica `Gleb Wataghin', Universidade Estadual de Campinas, 13083-859, Campinas, S\~{a}o Paulo, Brazil}
\ead{$^*$akramt1@umbc.edu}

\begin{abstract} 
It is an established fact that quantum coherences have thermodynamic value. The natural question arises, whether other genuine quantum properties such as entanglement can also be exploited to extract thermodynamic work. In the present analysis, we show that the ergotropy can be expressed as a function of the quantum mutual information, which demonstrates the contributions to the extractable work from classical and quantum correlations. More specifically, we analyze bipartite quantum systems with locally thermal states, such that the only contribution to the ergotropy originates in the correlations. Our findings are illustrated for a two-qubit system collectively coupled to a thermal bath.
\end{abstract}

\section{Introduction}\label{intro}

What is a resource in thermodynamics? From the inception of the theory, the question appears rather simple to answer -- namely a thermodynamic resource is any energy that can be extracted from, e.g., a heat or particle reservoir, and transformed into work~\cite{callen1998thermodynamics}. However, it has been debated since essentially the beginnings of thermodynamics if and to what extent \emph{information} can also be considered a resource~\cite{Leff2014,info1,info2,Zurek1989}.  Remarkably, the \emph{thermodynamics of information}~\cite{Parrondo2015,Wolpert2019} was fully established only rather recently, which was spurred by developing the stochastic thermodynamics with feedback~\cite{Sagawa2008,Sagawa2010} and by proposing the notion of information reservoirs~\cite{Deffner2013PRX}.

Evidently, describing information in quantum systems is more subtle, and hence also the thermodynamics of quantum information requires more thorough analyses~\cite{JPA_Goold,DeffnerCampbellBook}. This area of research has received a boost by the realization that quantum thermodynamics contributes profoundly to the development of new generation quantum technologies~\cite{Dowling_Milburn,PRXQuantum_Deutsch}. Among these emerging technologies, in particular quantum thermal machines~\cite{TJP_Asli,arXiv_Bhattacharjee,CP_Mark,arXiv_Latune,arXiv_Mukherjee,EPJST_Ghosh} and quantum information engines~\cite{Quan2006,Deffner2013PRE,Strasberg2017,Stevens2019,ashrafi2020szilard}, aka quantum computers~\cite{nielsen2002quantum} necessitate a comprehensive study of quantum information as a thermodynamic resource.  In this context, it is important to realize that from a thermodynamic perspective quantum information as quantified by the von Neumann entropy is not the only notion of information to be considered.  Rather,  understanding the contribution of information encoded in marginals~\cite{Polkovnikov2011,Deffner2020PRR,Deffner2021qubit}, and in particular the thermodynamic value of genuine quantum correlations~\cite{Chapman2015,Safranek2018,Entropy_Ceren} is instrumental.

Therefore, the present analysis specifically focuses on the role that correlations (quantum or classical) play in the maximum extractable work, i.e.,  the \emph{ergotropy}.  In the literature, this problem was partially addressed from different perspectives, by relating work extraction to either coherences~\cite{francica2020quantum,NJP_Korzekwa,PRE_Baris,PLA_Giacomo} or correlations~\cite{PRA_Funo,npjQI_Francica,Entropy_Mauro,PRL_Manzano,PRX_Llobet,PRE_Fusco,correlation1,correlation2,correlation3,add1,add2}, for a wide variety of scenarios. Nevertheless, all previous studies either rely on correlations with an ancillary system together with a feedback mechanism~\cite{PRA_Funo,npjQI_Francica,Entropy_Mauro,PRL_Manzano}, are restricted to specific dynamical models~\cite{PRE_Baris,PRE_Fusco}, or establish a connection between extractable work and the average energy for correlated states~\cite{PRX_Llobet}. To the best of our knowledge, a general mathematical relationship between the correlations among the constituents of a quantum system and its work content is still lacking. 

To address this void in our understanding of the thermodynamics of correlations, we prove a direct relationship between the ergotropy and the quantum mutual information in a bipartite, and locally thermal, quantum state.  The locally thermal states are a judicious choice such that the only resources available for work extraction are total bipartite correlations. Furthermore, we derive an equality relating the ergotropy to both the quantum mutual information and the bound ergotropy~\cite{bound}. We illustrate and analyze these general results in the specific context of an array of qubits collectively coupled to a thermal bath at a finite temperature. Finally, we upper bound the average power that can be extracted from an arbitrary quantum state. Our results explicitly quantify the role of correlations in the process of work extraction by means of cyclic and unitary processes, during which the system of interest is isolated and the only available resources are in the form of bipartite correlations.

\section{Ergotropy and the process of work extraction}\label{prelim}

Ergotropy is the maximum amount of work that can be extracted from a quantum system by means of cyclic and unitary operations~\cite{EPL_Allahverdyan}. From a thermodynamic standpoint,  work is then simply given by the change of internal energy, since the corresponding process is unitary, or in other words thermally isolated. Therefore, we consider the dynamics of a system governed by the Hamiltonian $H_{total}=H+\Gamma(t)$, where $H$ denotes the self-Hamiltonian of the system and $\Gamma(t)$ is a time-dependent coupling term responsible for the extraction of work during time $\tau$. Note that by construction $\Gamma(t)$  fulfills $\Gamma(0)=0$ and $\Gamma(\tau)=0$,  such that the operation is cyclic with respect to $H$.

Now, consider a quantum system described by $H=\sum_{i=1}^{d}\varepsilon_{i}\left|\varepsilon_{i}\right\rangle\left\langle\varepsilon_{i}\right|$  and quantum state $\rho=\sum_{j=1}^{d} r_{j}\left|r_{j}\right\rangle\left\langle r_{j}\right|$, such that $\varepsilon_{i} \le \varepsilon_{i+1}$ and $r_{j} \geq r_{j+1}$. The ergotropy is then calculated by performing an optimization over all possible unitary operations to achieve a final state that has the minimum average energy with respect to $H$,
\begin{equation}
    \mathcal{E}(\rho)=\trace{H\rho}-\miU{\trace{H U\rho U^{\dag}}}=\trace{H(\rho-P_{\rho})},
    \label{ergo}
\end{equation}
where $P_{\rho}\equiv\sum_{k} r_{k}\ket{\varepsilon_{k}}\bra{\varepsilon_{k}}$ is called the \emph{passive state}~\cite{pusz1978passive}.  By plugging the explicit form of $P_{\rho}$ in the equation above, we obtain the well-known expression~\cite{EPL_Allahverdyan}
\begin{equation}
\mathcal{E}(\rho)=\sum_{i, j} r_{j}\varepsilon_{i}\left(|\bra{r_j} \varepsilon_i \rangle|^2-\delta_{ij}\right).
\label{ergo2}
\end{equation}

The specific unitary $U$ that takes an arbitrary state $\rho$ to its corresponding passive state $P_{\rho}$ is given by $U=\sum_k\ket{\varepsilon_k}\bra{r_k}$. Hence, we have a general form for the potential $\Gamma(t)$, that generates the desired unitary operation, such that~\cite{EPL_Allahverdyan}
\begin{equation}
\Gamma(t)=\dot{\varphi}(t)\, \ex{-i H t / \hbar}\, \Lambda\, \ex{i H t / \hbar},
\label{pot}
\end{equation}
where $\varphi(0)=\dot{\varphi}(0)=\dot{\varphi}(\tau)=0$, $\varphi(\tau)=\tau$, and in the interaction picture $U_{I}(\tau) \equiv \ex{i H \tau / \hbar}\, U=\ex{-i \Lambda \tau / \hbar}$. Note that the freedom to choose $\varphi$ implies that the potential $\Gamma(t)$ is not unique. Finally, it is interesting to note that with the exception of thermal states, it is possible to extract work from multiple copies of passive states by processing them collectively~\cite{bound}. For this reason, thermal states are also referred to as \emph{completely passive states}~\cite{pusz1978passive}.

\section{Extractable work from correlations}\label{main}
\subsection{Ergotropy and the mutual information}
\label{seca}

For the following analysis, we consider a quantum system $\mc{S}$, that can be separated into two partitions $A$ and $B$. For such scenarios, we now prove a direct relationship between the ergotropy and the mutual information,
\begin{equation}
\mc{I}(A:B)\equiv S(\rho_A)+S(\rho_B)-S(\rho),
\label{mutual}
\end{equation}
where $S(\rho_i)=-\trace{\rho_i \ln(\rho_i)}$ denotes the von Neumann entropy, and $\rho$ is the quantum state of $\mc{S}$.

%%%%% Fig 1 %%%%%
\begin{figure}
\centering
\includegraphics[width=0.9\textwidth]{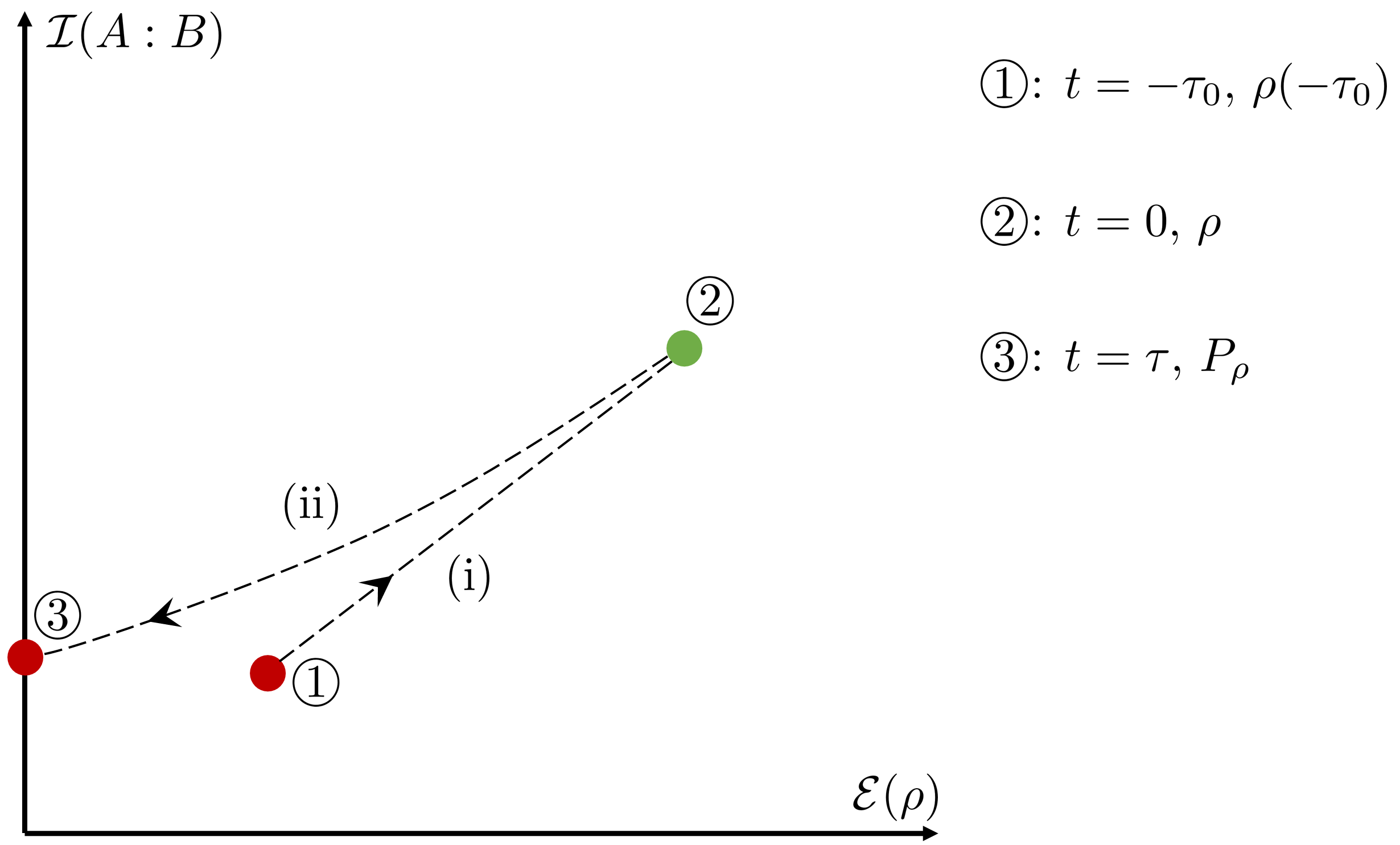}
\caption{Sketch of the successive processes (i) and (ii), where the build up of correlations between the partitions $A$ and $B$ (process (i)) through either open or closed dynamics enables work extraction by means of unitary evolution (process (ii)).}
\label{illustration}
\end{figure}
%%%% %%%%

Our goal is now to assess how much work can be extracted from the quantum correlations between $A$ and $B$. To this end, we consider a two-stroke operation on $\mc{S}$: (i) correlations are built-up and $\mc{S}$ is driven into thermal (passive) states in $A$ and $B$; and (ii) work is extracted under a cyclic, unitary operation on $\mc{S}$, for which we can compute the ergotropy. 

More specifically,  consider the following situation: (i) the quantum system $\mc{S}$ evolves from $t=-\tau_0$ to $t=0^-$ under
\begin{equation}
H(t)=H_A\otimes \mathds{I}_B+\mathds{I}_A\otimes H_B+H_I(t),
\end{equation}
where $H_I(t)$ contains interactions among the subsystems and their interactions with the environment. We assume that at $t=0$ the interaction Hamiltonian is negligible compared to the self-Hamiltonians of $A$ and $B$, i.e., $H(0)=H_A\otimes \mathds{I}_B+\mathds{I}_A\otimes H_B \equiv H$. This part can be viewed as a state preparation stage in which one tries to obtain a globally active state by means of creating and/or sustaining correlations between the bipartitions. We would like to highlight that there are no restrictions on the nature of the dynamics here, i.e., it can be open or closed. As a result, we evolve an initially arbitrary density matrix $\rho(-\tau_0)$ to a steady state density matrix $\rho$ such that both $\rho_A=\ptr{B}{\rho}=\ex{-\beta H_A}/Z_A$ and $\rho_B=\ptr{A}{\rho}=\ex{-\beta H_B}/Z_B$ are thermal states at inverse temperature $\beta$.

Then, (ii) work is extracted during time $\tau$ by unitary means. The latter implies that any possible interaction with the surroundings must be cut off, i.e., the system is isolated. As outlined above, the total Hamiltonian reads,
\begin{equation}
H(t)=H_A\otimes \mathds{I}_B+\mathds{I}_A\otimes H_B+\Gamma(t),
\end{equation}
where $\Gamma(0)=\Gamma(\tau)=0$. It is then a simple exercise to show that the ergotropy $\mc{E}(\rho)$ can be written as
\begin{equation}
\beta\,\mathcal{E}(\rho) =\mathcal{I}(A:B)-D(P_{\rho}||\rho_{A}\otimes \rho_{B}),
\label{eq:mainresult}
\end{equation}
where $D(\rho||\sigma)=\trace{\rho\ln(\rho)-\rho\ln(\sigma)}$ is the relative entropy, between the states $\rho$ and $\sigma$~\cite{nielsen2002quantum}.  Note that $\rho_{A}$ and $\rho_{B}$ and $\mc{I}(A:B)$ are evaluated at $t=0$ (see Fig.~\ref{illustration}). 

Equation~\eqref{eq:mainresult} constitutes our main result and shows that there are two additive contributions to the ergotropy of locally thermal states. While the mutual information between $A$ and $B$ is the main resource for the finite ergotropy, the distance between $P_{\rho}$ and the tensor product of local states, at the beginning of the work extraction process, reduces the amount of ergotropy, owing to the non-negativity of the relative entropy. This suggests that the \emph{optimal} work extraction process erases all correlations (quantum and classical) between the subsystems at the end of the process. However, in general this is not possible due to the fact that the entropies of the states $P_{\rho}$ and $\rho_{A}\otimes \rho_{B}$ can be different in magnitude\footnote{Equation~\eqref{eq:mainresult} can be equivalently obtained using the fact that for any two different density matrices $\rho_1$ and $\rho_2$, with $S(\rho_1)=S(\rho_2)$, we have
\begin{equation}
\beta\,\trace{(\rho_1-\rho_2)H}=D(\rho_1||\rho_{\beta})-D(\rho_2||\rho_{\beta}),
\label{eql1}
\end{equation}
where $\rho_{\beta}=\exp(-\beta H)/Z$ is the global thermal state.}. To further elaborate, there exists an interplay/trade-off between the mutual information $\mc{I}(A:B)$ and the Kullback-Leibler divergence $D(P_{\rho} \| \rho_A \otimes \rho_B)$. Maximizing the mutual information between the partitions does not guarantee higher ergotropy since the two terms in~\myeqref{eq:mainresult} can cancel each other and result in zero ergotropy for finite amount of correlations. Therefore, \myeqref{eq:mainresult} conveys that both the mutual information and $D(P_{\rho} \| \rho_A \otimes \rho_B)$ play an important role in the work extracted through cyclic and unitary operations.

Additionally, it is important to note that the Kullback-Leibler divergence $D(P_{\rho}||\rho_A\otimes\rho_B)$ is a measure of divergence, or the overlap between two density matrices~\cite{nielsen2002quantum} ($P_{\rho}$ and $\rho_A\otimes\rho_B$). From a thermodynamic standpoint, since we are dealing with locally thermal states, the quantity $D(P_{\rho}||\rho_A\otimes\rho_B)/\beta$ can be interpreted as the exergy, or the available energy, of the system in a relaxation process~\cite{kull1,kull2} (a process that would take the density matrix $P_{\rho}$ to $\rho_A\otimes\rho_B$). 

It is also interesting to note that~\myeqref{eq:mainresult} implies an ``inverse Landauer'' inequality for correlation,
\begin{equation}
\beta\,\mathcal{E}(\rho) \leq \mathcal{I}(A:B).
\label{ineq1}
\end{equation}
The maximum amount of work that can be extracted from a closed, bipartite quantum system with locally thermal states, is given by the mutual information between its subsystems. The bound is saturated if and only if $P_{\rho}=\rho_{A}\otimes \rho_{B}$.

Finally, it is interesting to see how the present findings relate to established work extraction schemes,  for which $\mc{S}$ is in contact with a heat bath. In general, the nonunitary schemes, in which the system is weakly coupled to the bath and the process is isothermal, outperform the unitary procedure since the presence of the bath lifts the constant entropy restriction on the system. In this case, the maximum amount of extractable work $\mathcal{E}_{\beta}(\rho)$ is determined by the difference in the non-equilibrium free energies of the state at hand and the thermal state, $\mathcal{E}_{\beta}(\sigma)=F_{\beta}(\sigma)-F_{\beta}(\sigma_{\beta})$, where $F_{\beta}(\sigma)=\trace{H\sigma}-1/\beta\,S(\sigma)$~\cite{Bergmann1955,EPL_Esposito,NatPhys_Parrondo,PRL_Manzano}. For a general bipartite system, it is then possible to write~\cite{PRL_Manzano}
\begin{equation}
\begin{split}
\mathcal{E}_{\beta}(\rho)&=F_{\beta}(\rho)-F_{\beta}(\rho_{A,\beta}\otimes\rho_{B,\beta}),\\ 
&=\mathcal{E}_{\beta}(\rho_A)+\mathcal{E}_{\beta}(\rho_B)+1/\beta\,  \mathcal{I}(A:B). 
\end{split}
\label{eq:Manzano}
\end{equation}
In our case, we have $\mathcal{E}_{\beta}(\rho_A)=\mathcal{E}_{\beta}(\rho_B)=0$.  Comparing this with \myeqref{eq:mainresult} we can immediately identify the difference between the maximum extractable work in unitary and non-unitary approaches as $D(P_{\rho}||\rho_{A}\otimes \rho_{B})$. Shortly, we will see the present work extraction scheme (for process (ii)) can be further improved,  when we utilize multiple copies of the same quantum state.

\subsection{Bound ergotropy and multipartite correlations}\label{sec:bound}

For a given entropy, the quantum state with the minimum average energy is the thermal state. For states $P_{\rho}$, that are passive, but not thermal, i.e., not completely passive, the remaining energy in $P_{\rho}$ can be accessed by implementing a secondary process. It has been shown that this secondary process can be constructed through multiple copies of $P_{\rho}$, and the additionally extractable work has been dubbed \textit{bound ergotropy} $\mc{E}_b(\rho)$~\cite{bound}. It can be written as
\begin{equation}
\label{eq:bound}
\mc{E}_b(\rho)= \trace{(P_{\rho}-P^{\text{th}}_{\rho}) H}\geq 0,
\end{equation}
where $P^{\text{th}}_{\rho}$ is the thermal state corresponding to $P_{\rho}$ such that $S(P_{\rho})=S(P^{\text{th}}_{\rho})$. In other words, the bound ergotropy is the amount of additional ergotropy that can be extracted from $N$ copies of the system per copy in the limit of $N\rightarrow\infty$. Exploiting Eqs.~\eqref{eq:mainresult} and \eqref{eql1} we further have
\begin{equation}
\beta\left(\mathcal{E}(\rho)+\mathcal{E}_b(\rho)\right) =\mathcal{I}(A:B)-D(P^{\text{th}}_{\rho}||\rho_{A}\otimes \rho_{B})\,.
\label{attempt3}
\end{equation}
Again noting that the relative entropy is non-negative, we can write 
\begin{equation}
\beta\,\left(\mathcal{E}(\rho)+\mathcal{E}_b(\rho)\right) \leq \mathcal{I}(A:B).
\label{ineq2}
\end{equation}
Equation~\eqref{ineq2} is a tighter version of the inverse Landauer's principle \eqref{eq:mainresult}, and it is in fact tight. Below in Sec.~\ref{examples}, we will show that the bound can be saturated for qubits collectively coupled to a thermal bath.

Moreover, it is interesting to note that the ergotropy is a non-extensive quantity. Namely, it is easy to see from the strong subadditivity of the von Neumann entropy that the global ergotropy, $\mathcal{E}_{G}(\rho)$, for $N$ copies of  $\rho$ is greater or equal to $N \mc{E}(\rho)$. Hence, we have
\begin{equation}
\beta\,\mathcal{E}_{G}(\rho) \equiv \beta\,N\left(\mathcal{E}(\rho)+\mathcal{E}_b(\rho)\right) \leq N\,\mathcal{I}(A:B)\,.
\label{global}
\end{equation}
\begin{comment}
which is a consequence of the strong subadditivity of the von Neumann entropy.
\end{comment}
Note, however, that the bound ergotropy can be extracted by acting globally on the $N$ copies of $\rho$ (since $P_{\rho}$ is not completely passive).  Inequality~\eqref{global} is saturated if and only if the thermal state can be expressed as follows $\rho_A \otimes \rho_B=P^{\text{th}}_{\rho}$.

We conclude this section by noting that Eqs.~\eqref{eq:mainresult} and \eqref{global} can be readily generalized to multipartite correlations. Using \myeqref{eql1}, and assuming that $\mc{S}$ is composed of ``$k$'' thermal states correlated with each other, we obtain
\begin{equation}
\beta\,\mathcal{E}(\rho) \leq  \mathcal{I}(A_1:A_2:...:A_k),
\label{multi1}
\end{equation}
and
\begin{equation}
\beta\,\left(\mathcal{E}(\rho)+\mathcal{E}_b(\rho)\right) \leq \,\mathcal{I}(A_1:A_2:...:A_k),
\label{multi2}
\end{equation}
where $\mc{I}(A_1:A_2:...:A_k)=D(\rho_{A_1...A_k}||\rho_{A_1}\otimes ... \otimes \rho_{A_k})$ is the multipartite mutual information between the $k$ partitions of $\mc{S}$~\cite{multipart}.

\subsection{Ergotropy and quantum discord}

We continue with a closer analysis of the nature of the correlations. It has been established that the quantum mutual information quantifies the amount of total correlations between two parties $A$ and $B$~\cite{Henderson_2001,discord}. In particular, this means that $\mathcal{I}(A:B)$ is comprised of classical \emph{and} quantum correlations.  An accepted measure for the purely quantum contribution is the \emph{quantum discord}~\cite{discord}, which is typically written as
\begin{equation}
\mf{D}(A:B)=\mathcal{I}(A:B)-J(A:B),
\end{equation}
where $J(A:B)$ is the Holevo information~\cite{nielsen2002quantum} that quantifies the maximum amount of classical information that is determined from optimized generalized measurements on $B$. Hence,  $J(A:B)$ quantifies the maximal, classical information that $B$ can carry about $A$. Therefore,  $\mf{D}(A:B)$ is the genuinely quantum information. Using the definition of the quantum discord, Eq.~\eqref{eq:mainresult} can be written as
\begin{equation}
\beta\,\mathcal{E}(\rho)=\mf{D}(A:B)+J(A:B)- D(P_{\rho}||\rho_{A}\otimes \rho_{B}),
\label{attempt3a}
\end{equation}
which highlights the interplay of classical and quantum correlations. It is then instructive to consider under what conditions and for what states $\rho_0$ the ergotropy vanishes, $\mathcal{E}(\rho_0)=0$. For such $\rho_0$ we have
\begin{equation}
\mf{D}(A:B)+J(A:B)= D(P_{\rho_0}||\rho_{A}\otimes \rho_{B})\,.
\end{equation}
We immediately conclude that the existence of correlations (quantum or classical) is necessary, but not sufficient, for a non-zero ergotropy in locally thermal states.

\subsection{Ergotropy and system-environment correlations}

As alluded to above,  $\mathcal{E}(\rho)$ is a lower bound on the work that can be extracted by non-unitary operations on $\mc{S}$,
\begin{equation}
\mathcal{E}(\rho) \leq \mathcal{E}_{\beta}(\rho).
\end{equation}
which follows from exploiting correlations between $\mc{S}$ and its environment \cite{PRA_Funo,PRL_Manzano,Entropy_Mauro,npjQI_Francica}.  Additionally, Eq.~\eqref{eq:mainresult} can also be written as \cite{touil2020information},
\begin{equation}
\beta\,\mathcal{E}(\rho)=\Delta S_A+\Delta S_B-\mc{I}(\mc{S}:E)-\beta \la Q \ra-D(P_{\rho}||\rho_{A}\otimes \rho_{B}),
\label{SE}
\end{equation}
where $\la Q \ra$ is the heat exchanged between system and environment, and $\mc{I}(\mc{S}:E)$ quantifies the buildup of correlations.  Equation~\eqref{SE} demonstrates that system-environment correlations and the dissipated heat diminish the amount of work that can be extracted by unitary processes.  This insight is complementary to what has been shown in the literature \cite{lit3,lit5,npjQI_Francica}, and emphasizes our different approach that focuses on work extraction by means of cyclic and unitary operations on a state that is locally completely passive.\footnote{As a side note, it is interesting to consider our result, in~\myeqref{SE}, in context of work extraction from information scrambling \cite{scr1,scr2,scr3,scr4,QST_Akram,zanardi2020information,touil2020information}.  However, a thorough analysis of extracting work from scrambled states, or even from quantum chaos, is beyond the scope of the present work, see also Refs.~\cite{PRL_SYK,JHEP_SYK}.}

\section{Illustrative case study: ergotropy from X-states}\label{examples}

After having established the conceptual framework, the remainder of the analysis is dedicated to an instructive case study. In particular, we elucidate the conditions and physical mechanisms that lead to locally thermal states.

\subsection{Two-qubit systems and X-states}

Consider now that $\mc{S}$ is comprised of two qubits, which are initially prepared in an arbitrary quantum state.  Note that working with qubits makes the analysis particularly simple, since any diagonal qubit state can be described by a thermal state at an effective inverse temperature $\beta$. A straightforward, though not the most general, way to then have locally thermal states is if $\mc{S}$ relaxes into an X-state
\begin{equation}\label{xshape}
\rho(t)=
\begin{pmatrix}
\rho_{11} & 0 & 0 & \rho_{14} \\
0 & \rho_{22} & \rho_{23} & 0 \\
0 & \rho_{23}^* & \rho_{33} & 0 \\
\rho_{14}^* & 0 & 0 & \rho_{44} \\
\end{pmatrix}\,.
\end{equation}
Such states have been widely studied in the literature~\cite{x_states}, as they can be found in generalized Pauli channels~\cite{xshape} or in collective dephasing models for two-qubit systems~\cite{NJP_Carnio}. Moreover, the ground state of one-dimensional spin chains that are invariant under translations, and parity transformations, i.e., exhibit $\mathbb{Z}_2$ symmetry, also have reduced bipartite density matrices in X-shape~\cite{RMP_Amico,Sarandy,Entropy_Baris}. For a more detailed discussion of X-states we refer to the literature~\cite{x_states}.

\subsection{A physical model: Collective dissipation}

We continue with an even more specific scenario, and now consider two interacting qubits, which are assumed to behave as point-like dipoles with identical dipole moments, and which are immersed in a finite-temperature environment composed of thermal photons resonant with the qubits' transition frequency. The master equation governing the dynamics is given by~\cite{PRA_Angsar,PRA_Latune,Gross1982,Stephen,Lehmberg,JPhysB_Damanet}
\begin{equation}
\frac{\pd \rho}{\pd t} = -\frac{i}{\hbar}[(H_0+H_d), \rho]+\mathcal{D}_-(\rho)+\mathcal{D}_+(\rho)= \mathcal{L}(\rho),
\end{equation}
where $H_0\!=\!\hbar\omega(\sigma_1^+\sigma_1^-+\sigma_2^+\sigma_2^-)$ and $H_d\!=\!\hbar f(\sigma_1^+\sigma_2^-+\sigma_2^+\sigma_1^-)$ are the self-Hamiltonian of the whole system and the interaction Hamiltonian between the qubits, respectively. As usual, $\sigma_i^+\!=\!|e_i\rangle\langle g_i|$ ($\sigma_i^-=(\sigma_i^+)^{\dagger}$) is the raising (lowering) operator with $i \in \{1, 2\}$ denoting the qubit label. The explicit form of the dissipators describing the interaction with the bath reads
\begin{equation}
\begin{split}
\mathcal{D}_-(\rho) &=\sum\limits_{i,j=1}^2\gamma_{ij}\,(\bar{n}+1)(\sigma_j^-\rho\sigma_i^+-\frac{1}{2}\{\sigma_i^+\sigma_j^-, \rho\}), \\ 
\mathcal{D}_+(\rho) &=\sum\limits_{i,j=1}^2\gamma_{ij}\,\bar{n}(\sigma_j^+\rho\sigma_i^--\frac{1}{2}\{\sigma_i^-\sigma_j^+, \rho\}).
\end{split}
\end{equation}
Here, $\bar{n}\!=\![\exp(\beta_e\omega)-1]^{-1}$ is the mean number of photons at the temperature of the environment $\beta_e$, and $\gamma_{ij}$ are the spontaneous decay rates. Without loss of generality, we assume $\gamma_{ij}=\gamma_{ji}$.

Depending on the spatial configuration of the system qubits, the model above has two distinct limits. If the qubits are well separated, i.e., the distance between them is much larger than the wavelength of the thermal photons in the environment, they individually couple to the bath and their steady-state is simply described by the Gibbs state, at the bath temperature, $\rho_{\beta_e}=\exp(-\beta_e H_0)/Z$. Naturally, such a steady-state is completely passive.  On the other hand, when the qubits are closely packed, such that the separation between them is much smaller than the wavelength of the thermal photons, they collectively couple to the environment. In this regime, the qubits become indistinguishable due to the impossibility of resolving which qubit absorbed or emitted a photon. Thus,  their steady-state admits coherences in the energy eigenbasis~\cite{PRA_Angsar,PRA_Latune} and has the following form~\cite{PRA_Latune}
\begin{align}\label{ss_twoqubit}
\rho_{ss}&= (1-c)\ket{\psi_-}\bra{\psi_-} \\ \nonumber
& +cZ_+^{-1}\left(\beta_e\right)\left(\ex{-2\omega\beta_e}\ket{\psi_{ee}}\bra{\psi_{ee}}+\ex{-\omega\beta_e}\ket{\psi_+}\bra{\psi_+}+\ket{\psi_{gg}}\bra{\psi_{gg}}\right).
\end{align}
Here, $\ket{\psi_{gg}}\!=\!\ket{gg}$, $\ket{\psi_{ee}}\!=\!\ket{ee}$, $\ket{\psi_{\pm}}\!=\!\ket{ge}\pm\ket{eg}/\sqrt{2}$, $c\!=\!\bra{\psi_{gg}}\rho_0\ket{\psi_{gg}}+\bra{\psi_{ee}}\rho_0\ket{\psi_{ee}}+\bra{\psi_+}\rho_0\ket{\psi_+}$, and $Z_+\left(\beta_e\right)\!=\!1+\ex{-\omega\beta_e}+\ex{-2\omega\beta_e}$. 

Consequently, the steady-state preserves some information about the initial state of the dynamics through the parameter ``$c$'', and thus it is not unique. Only recently, it has been shown that such states possess a finite amount of ergotropy~\cite{PRE_Baris} for a wide range of initial states and environment temperatures. Further, Eq.~\eqref{ss_twoqubit} is also in the X-shape form \eqref{xshape} with $\rho_{14}=0$. Note, however, that the local temperature of the qubits, $\beta$, is generally different from the bath temperature $\beta_e$. In fact, it is a simple exercise to show that
\begin{equation}
\beta=\frac{1}{\omega}\ln\left[\frac{1+2\cosh(\beta_e\omega)+2c\sinh(\beta_e\omega)}{1+2\cosh(\beta_e\omega)-2c\sinh(\beta_e\omega)}\right].
\end{equation}

As we have seen above \eqref{eq:mainresult}, to maximize the ergotropy the target passive state needs to be as close as possible to the product state $\rho_{A}\otimes \rho_{B}$.  However, achieving such a passive state through any unitary process is generally not possible since the states $P_{\rho}$ and $\rho_{A}\otimes \rho_{B}$ have different values of entropy. Despite this fact, it is still possible to further process the passive state $P_{\rho}$ provided that it is not a completely passive state, and we have access to, and ability to act on, multiple copies of it. To this end, we are interested in the bound ergotropy $\mc{E}_b(\rho)$ \eqref{eq:bound}, which also can be determined analytically.

\begin{figure}[t]
{\bf (a)} \hskip0.33\columnwidth {\bf (b)}\hskip0.33\columnwidth {\bf (c)}\\
\includegraphics[width=0.33\columnwidth]{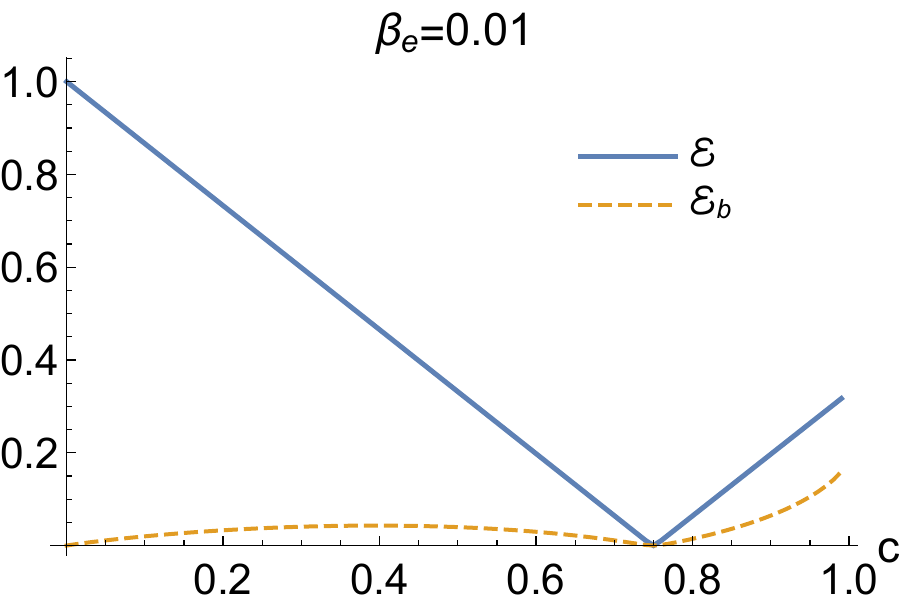}\includegraphics[width=0.33\columnwidth]{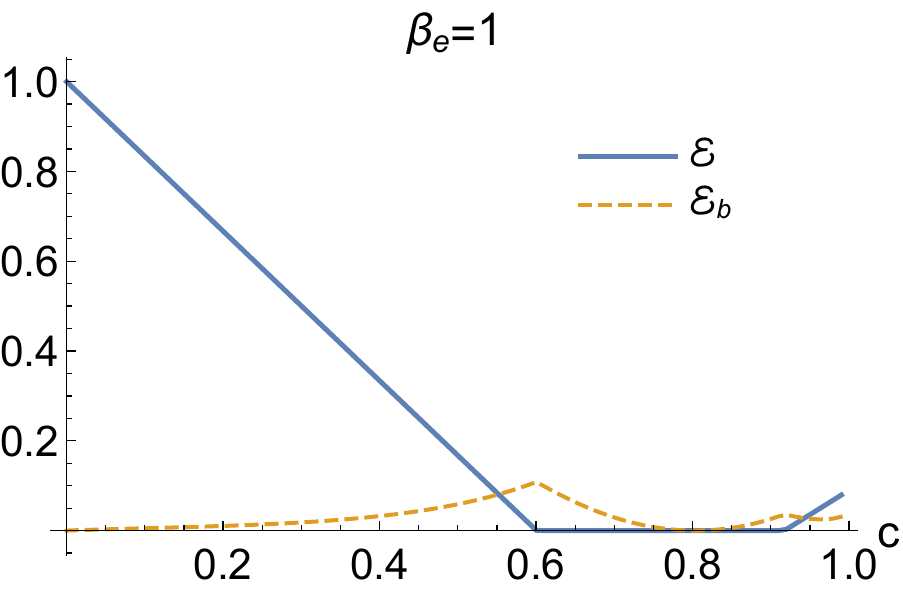}\includegraphics[width=0.33\columnwidth]{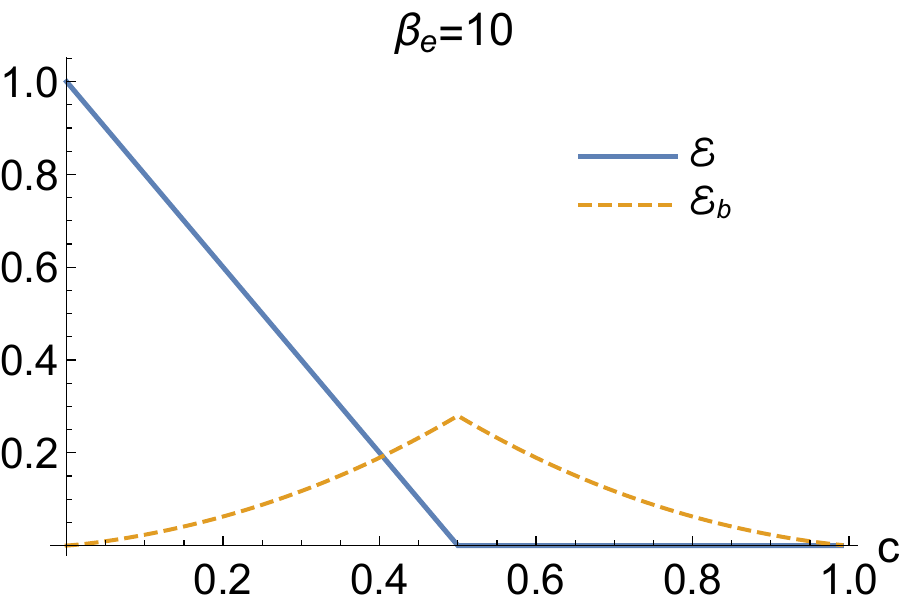}
\caption{Ergotropy and bound ergotropy as a function of the initial state parameter $c$ for $\beta_e=0.01$ {\bf (a)}, $\beta_e=1$  {\bf (b)} and $\beta_e=10$ {\bf (c)}.}
\label{fig:EandEb}
\end{figure}

In Fig.~\ref{fig:EandEb}, we plot $\mc{E}(\rho)$ and $\mc{E}_b(\rho)$ as a function of $c$, that is, for different initial states of the open system dynamics. Note that the ergotropy is evaluated using \myeqref{eq:mainresult}. In contrast to the ergotropy, which bears a finite value for nearly all values of $c$ at high environment temperatures (low $\beta_e$), the overall magnitude of bound ergotropy is the largest at low environment temperatures (high $\beta_e$). However, the amount of the bound ergotropy does not represent a significant amount as compared to the ergotropy. A two-qubit system with a self-Hamiltonian of $H_0$ has an energy gap of $2\hbar\omega$, which corresponds to the highest amount of ergotropy that one can get out of this system if both qubits were to be in their excited states. Comparing $\mathcal{E}_b$ with the maximum we observe for the steady-states of the considered model and with the absolute maximum of $2\hbar\omega$, we conclude that $\mathcal{E}_b$ is reasonably smaller than both of these quantities, and the difference is more pronounced for the latter case. This result can be expected considering that the bound ergotropy is actually the amount of extractable work from a state that is already passive, and can only be accessed by acting on multiple copies. 

The zero of the ergotropy corresponds to situations for which $\mc{S}$ is prepared in a thermal state at inverse temperature $\beta_e$.  For such states it has actually been shown that the quantum system can not maintain any ergotropy in the long time limit \cite{PRE_Baris}. Curiously, this behavior is observed for a range of values of $c$, cf. Fig.~\ref{fig:EandEb}.

\begin{figure}[t]
{\bf (a)} \hskip0.33\columnwidth {\bf (b)}\hskip0.33\columnwidth {\bf (c)}\\
\includegraphics[width=0.33\columnwidth]{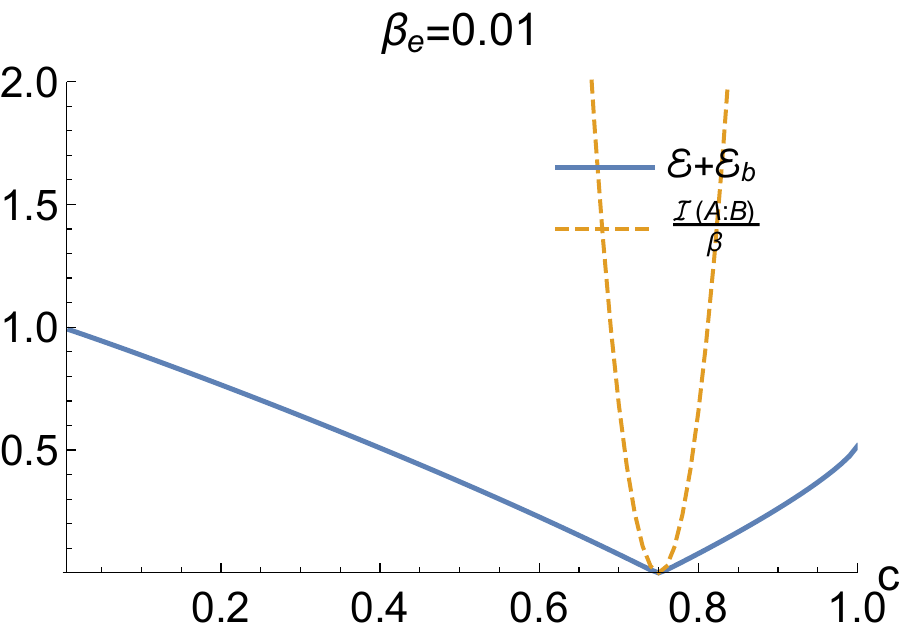}\includegraphics[width=0.33\columnwidth]{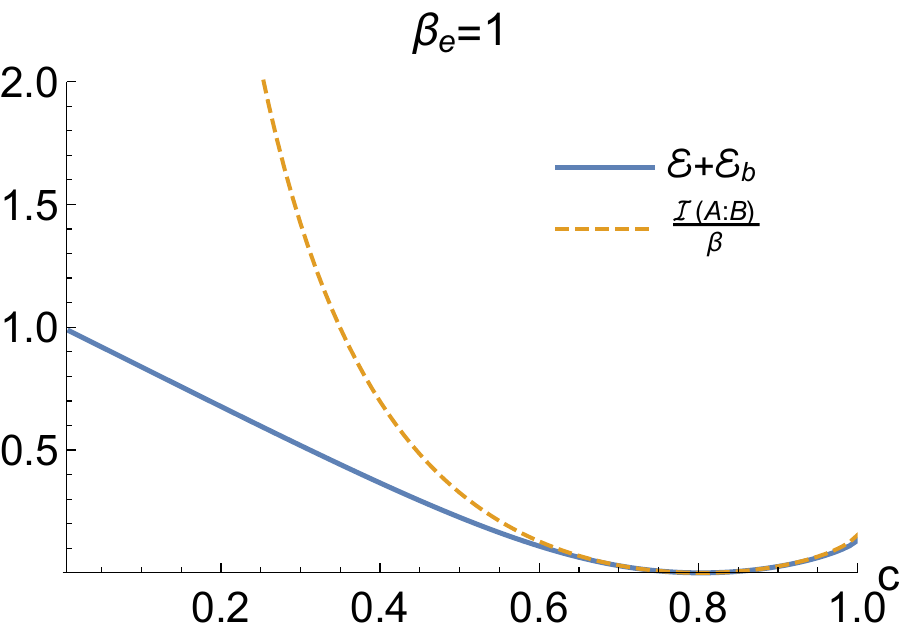}\includegraphics[width=0.33\columnwidth]{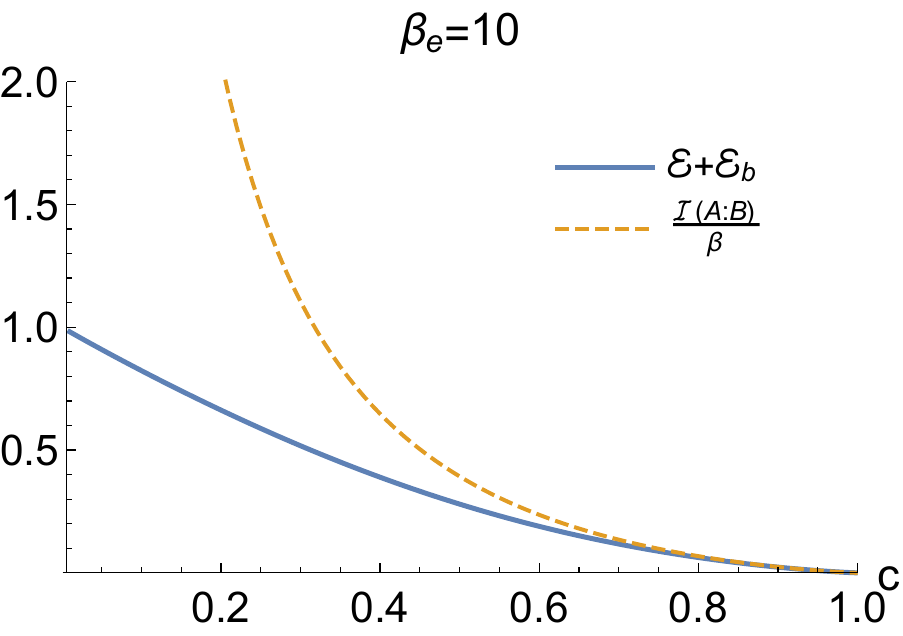}
\caption{Total ergotropy ($\mathcal{E}+\mathcal{E}_b$) and $\mathcal{I}(A:B)/\beta$ as a function of the initial state parameter $c$ for $\beta_e=0.01$ {\bf (a)}, $\beta_e=1$  {\bf (b)} and $\beta_e=10$ {\bf (c)}.}
\label{fig:totalEandMI}
\end{figure}

Finally, in Fig.~\ref{fig:totalEandMI}, we depict the global ergotropy, $\mathcal{E}_G=\mathcal{E}+\mathcal{E}_b$, together with $\mathcal{I}(A:B)/\beta$. We observe that the bound is far from tight for high environment temperatures as displayed in Fig.~\ref{fig:totalEandMI} {\bf (a)}. The only point where we have equality $\beta\mathcal{E}_G=\mathcal{I}(A:B)$ is when both quantities vanish at $c=0.75$. This value of the parameter $c$ corresponds to that of a thermal initial state at the considered environment temperature and hence neither ergotropy nor any form of correlation is generated as a result of the interaction with the bath. As we lower the temperature in Fig.~\ref{fig:totalEandMI} {\bf(b)} and {\bf (c)}, i.e., increase $\beta_e$, we observe that the bound is sharp for a wide range of initial states of the model. Therefore, it is interesting to observe that the ergotropy is aptly estimated by the mutual information, which is a lot simpler to compute for high-dimensional scenarios. Also note that the set of states for which we attain equality in \myeqref{ineq2} includes the ground initial state ($c=1$), which does not contain any quantum or classical correlations before the open system evolution. 

From a practical point of view, one problem that is critical in the extraction of the bound ergotropy is the preparation of multiple copies of a quantum state. In this sense, the present model for collective dissipation can be considered minimal in terms of the resources involved in the state preparation. An alternative approach, that is particularly useful in creating a number of copies of bipartite states, is a translationally invariant spin chain with $\mathbb{Z}_2$ symmetry~~\cite{RMP_Amico,Sarandy,Entropy_Baris}. As mentioned above, the ground states of such spin chains have reduced bipartite density matrices in the X-shape (especially see Sec. 3.3 of~\cite{Sarandy}), and are thus locally thermal. Although it is in general not an easy task to prepare a complex many-body system in its ground state, if achieved in a chain of $N$ spins (with $N$ as an even number), we end up with $N/2$ copies of the same X-shaped state.

\section{Bound on the average extractable power}

Finally, we would like to present few considerations on more practical matters. In all realistic scenarios, one is arguably more interested in the average power, rather than in the total amount of work that can be generated. Thus we now define
\begin{equation}
\mc{P}(\rho)\equiv\frac{\mc{E}(\rho)}{\tau} \ .
\label{power}
\end{equation}
The natural question arises whether the subtleties of quantum dynamics, in particular the quantum speed limit \cite{QST1}, pose additional constraints.  For unitary dynamics under driven Hamiltonians it has been shown that a generalized Mandelstam-Tamm bound takes the form \cite{Deffner2013JPA}
\begin{equation}
\tau \geq \frac{\hbar}{\Delta E_{\tau}} \mathcal{L}\left(\rho, P_\rho\right),
\label{QST}
\end{equation}
where $\mc{L}$ is the Bures angle \cite{bu1,bu2}
\begin{equation}
\mathcal{L}\left(\rho, P_\rho\right)=\arccos \left(\operatorname{tr}\left\{\sqrt{\sqrt{\rho} P_\rho \sqrt{\rho}}\right\}\right)\,.
\end{equation}
Further,  $\Delta E_{\tau}$ is the time-averaged variance of the Hamiltonian
\begin{equation}
\Delta E_{\tau} \equiv \frac{1}{\tau} \int_{0}^{\tau} dt\,  \Delta H_{t} =\frac{1}{\tau} \int_{0}^{\tau} dt\, \sqrt{\left\langle H(t)^{2}\right\rangle-\left\langle H(t)\right\rangle^{2}}.
\label{variance}
\end{equation}

While there are many formulations of the quantum speed limit, see Ref.~ \cite{QST1,Deffner2017NJP,Deffner2020PRR,o2020action} and references therein, this version is particularly convenient for the present purposes. In particular,  $\Delta E_{\tau}$ is upper bounded by using the \textit{finite energy constraint} \cite{bandwidth1,bandwidth2,SebastianViews} and the expression of $\Gamma(t)$ in~\myeqref{pot}.  We have,
\begin{equation}
(\exists \ \Omega>0) \ (\forall t \in [0, \ \tau]); \  \trace{H^{2}(t)} \leq \Omega^{2},
\end{equation}
which implies
\begin{equation}
(\exists \ \Omega>0) \ (\forall t \in [0, \ \tau]); \  \trace{\rho(t) H^{2}(t)} \leq \Omega^{2}.
\end{equation}
The above conditions state that the energy bandwidth of the Hamiltonian is uniformly bounded, simply implying that the physical resources at our disposal are limited.  Therefore, we can write
\begin{equation}
\label{eq:power}
\mc{P}(\rho) \leq \mc{I}(A:B) \frac{G(\Omega,\Lambda)}{\hbar\beta\, \mathcal{L}\left(\rho, P_\rho\right)}.
\end{equation}
Equation~\eqref{eq:power} gives an upper bound on the average extractable power that only depends on $\Omega$ (the upper bound on the bandwidth of the Hamiltonian $H(t)$), the time-independent potential $\Lambda$, the temperature $T$, the overlap between the initial state $\rho$ and the passive state $P_{\rho}$, and the mutual information. Note that this result is independent of the process duration $\tau$. We have,
\begin{equation}
G(\Omega,\Lambda)=\mi{\Omega,\sqrt{\left(\Omega +  \la \Lambda \ra_{0} + \trace{H_{AB}}\right)(\Omega - \la \Lambda \ra_{0})}},
\label{bound1}
\end{equation}
such that $\la\Lambda \ra_0=\trace{\rho\Lambda} $. Details of the derivation can be found in  \aref{a}. 

\section{Concluding remarks}\label{conclusion}
\subsection{Outlook and implications}
We conclude the analysis with a few remarks on the direct implications of our main results, and how they relate to the existing literature. Interestingly, it is straightforward to define a notion of \emph{efficiency} relating the amount of energy expended in process (i) to create correlations, and the useful work extracted during process (ii). The work of formation ``$W_f$'' quantifying the energetic cost associated with creating correlations was studied in Ref.~\cite{correlation1},  and it was found to be directly related to the multipartite mutual information. Extending this result to arbitrarily correlated states we can define the upper bound on efficiency as $\eta \equiv \mc{E}(\rho)/W_f$, see also Ref.~\cite{Niedenzu2018}. We leave the detailed analysis of this bound, and its relation to the mutual information, for future work. However, it is worth emphasizing already at this point that such an analysis will constitute an essential step towards a more complete thermodynamic description of quantum information processing.

Moreover, from a big picture perspective, we point out that the main difference between our work and what has been established in the literature~\cite{PRL_Manzano,PRX_Llobet} resides in the fact that we consider the relationship between the ergotropy (that is work extracted through cyclic and unitary processes) and the total correlations, as opposed to the case of work extraction in the presence of a heat bath. In Ref.~\cite{PRX_Llobet}, the authors considered locally thermal states and concluded  that  the  mutual information plays an important role in work extraction, in the presence of a heat bath. In our analysis, complementary to the result of Ref.~\cite{PRX_Llobet}, we showcase the seminal role of the mutual information in the ergotropy extracted from locally thermal states. Additionally, our result~(\ref{eq:mainresult}) can be easily generalized to arbitrary states where we have to account for additional terms in the ergotropy. Namely, it is a simple exercise to show that the ergotropy, for general mixed states $\rho$, reads
\begin{equation}
\beta \mc{E}(\rho)=\mc{I}(A:B)-D\left(P_{\rho} \| \rho_{\beta}\right)+\la\ln\left(\frac{\rho_A \otimes \rho_B}{\rho_{\beta}}\right)\ra_{\chi}+D(\rho_A \otimes \rho_B \| \rho_{\beta}),
\label{genres}
\end{equation}
such that $\chi=\rho-\rho_A \otimes \rho_B$ is the correlation matrix~\cite{corrmatrix1,corrmatrix2}, and $\la\ln\left(\frac{\rho_A \otimes \rho_B}{\rho_{\beta}}\right)\ra_{\chi}= \trace{\chi\ln\left(\frac{\rho_A \otimes \rho_B}{\rho_{\beta}}\right)}$~\footnote{More specifically, starting from the definition of the ergotropy, and for a general quantum state $\rho$ (such that its partitions are non-thermal), we get
\begin{equation}
\begin{split}
\beta \mc{E}(\rho)&=\beta \trace{(\rho-P_{\rho}) H},\\
&=D\left(\rho \| \rho_{\beta}\right)-D\left(P_{\rho} \| \rho_{\beta}\right),\\
&=\mc{I}(A:B)-D\left(P_{\rho} \| \rho_{\beta}\right)+\trace{\rho \ln\left(\frac{\rho_A \otimes \rho_B}{\rho_{\beta}}\right)}.\\
\end{split}
\end{equation}
Introducing the correlation matrix~\cite{corrmatrix1,corrmatrix2}, $\chi=\rho-\rho_A \otimes \rho_B$, results in the general equality presented in~\myeqref{genres}.}. Note that the last two terms in the above equality vanish in the case of locally thermal state, where we have $\rho_{\beta}=\rho_{A}\otimes \rho_{B}$, and we recover our main result in~\myeqref{eq:mainresult}.
\subsection{Summary}
In the present analysis, we derived a general equality relating ergotropy to the quantum mutual information between thermal partitions, $A$ and $B$, of a bipartite quantum system $\mc{S}$. Our conceptual considerations are based on two successive processes: we first create bipartite correlations as $\mc{S}$ is coupled to a thermal bath, then we decouple the system from the bath and extract the work from the resulting state of $\mc{S}$, through a cyclic and unitary process. For such scenarios, we proved a relationship between ergotropy, bound ergotropy, and the quantum mutual information. The resulting inequalities were demonstrated for the experimentally relevant scenario of an array of qubits collectively coupled to a thermal bath at a finite temperature.  Our results demonstrate that correlations shared within a quantum system in fact pertain thermodynamic value while being totally absent in its local states. Such a conclusion can prove to be useful in designing many-body quantum batteries, which are quantum systems utilized as a work reservoir, in a non-trivial setting. 

Finally, we sketched an analysis of the maximal average power that can be extracted from quantum correlations. However, due to the intricacies of a more thorough study, we leave bounds on the instantaneous power for future work.

\section*{Acknowledgements}
B. \c{C}. is supported by the BAGEP Award of the Science Academy, Research Fund of Bah\c{c}e\c{s}ehir University (BAUBAP) under project no: BAP.2019.02.03, and Technological Research Council of Turkey (TUBITAK) under Grant No. 121F246. S.D. acknowledges support from the U.S. National Science Foundation under Grant No. DMR-2010127. This research was supported by grant number FQXi-RFP-1808 from the Foundational Questions Institute and Fetzer Franklin Fund, a donor advised fund of Silicon Valley Community Foundation (SD).

\invisiblesection{Appendix}
\section*{Appendix A. Bounding the average  extractable power}
\label{a}
\renewcommand{\theequation}{A.\arabic{equation}}
\setcounter{equation}{0}  

In this appendix, we outline the derivation of Eq.~\eqref{eq:power}. To this end, we consider Hamiltonians fulfilling the finite energy constraint. For such situations we have
\begin{equation}
\begin{split}
\Delta E_{\tau}  &=\frac{1}{\tau} \int_{0}^{\tau}dt\,  \sqrt{\left\langle H(t)^{2}\right\rangle-\left\langle H(t)\right\rangle^{2}} ,\\
&\leq \frac{1}{\tau} \int_{0}^{\tau} dt\, \sqrt{\Omega^{2}-\left\langle H(t)\right\rangle^{2}} ,\\
&\leq \frac{1}{\tau} \sqrt{\int_{0}^{\tau} dt \left(\Omega-\left\langle H(t)\right\rangle\right)} \,\sqrt{\int_{0}^{\tau}dt\,  \left(\Omega+\left\langle H(t)\right\rangle\right)},\\
\label{try1}
\end{split}
\end{equation}
which follows from the  Cauchy–Schwarz inequality. Using the fact that $0 \leq 1/\tau\, \int_{0}^{\tau}dt\, \trace{ \rho(t) H_{AB} } \leq 1/\tau \int_{0}^{\tau} dt\, \trace{ H_{AB} } $ when the self-Hamiltonian's eigenvalues are greater or equal to zero, we obtain
\begin{equation}
 \Delta E_{\tau} \leq \sqrt{\Omega - \frac{1}{\tau} \int_{0}^{\tau} dt\, \left\langle \Gamma(t) \right\rangle}\,\sqrt{\Omega+ \frac{1}{\tau} \int_{0}^{\tau} dt \left\langle \Gamma(t) \right\rangle + \trace{H_{AB}} }.
\end{equation}
Furthermore, it can be shown that
\begin{equation}
\begin{split}
\frac{1}{\tau} \int_{0}^{\tau}dt \left\langle \Gamma(t) \right\rangle = \trace{\rho  \Lambda }\equiv  \la \Lambda \ra_{0}\,.
\end{split}
\end{equation}
In conclusion, the time-averaged variance is upper bounded by
\begin{equation}
\Delta E_{\tau} \leq \sqrt{\left(\Omega +  \la \Lambda \ra_{0} + \trace{H_{AB}}\right)(\Omega - \la \Lambda \ra_{0})}\,,
\end{equation}
from which we obtain Eq.~\eqref{eq:power}.

%\invisiblesection{References}
\section*{References}
\bibliographystyle{iopart-num}
\bibliography{ergo}

\end{document}